\documentclass[
    ,final 
  ,numberedheadings 
  ,cmfonts          
  ,float
  ,subfigure
  ]
  {aipproc}

\usepackage{amsmath}
\usepackage{subfigure}
\layoutstyle{6x9}

\newcommand{\ce}[1]{Eq.~(\ref{#1})}

\newcommand{\eqs}[1]{\begin{equation} \begin{split} #1\end{split} \end{equation} }

\newcommand{\ks}[1]{#1 \!\!\!\!\! \slash } 

\newcommand{\ga}{\gamma^5}

\newcommand{\ie}{{\it i.e.}}
\newcommand{\eg}{{\it e.g.}}
\newcommand{\etal}{{\it et al.}}

\renewcommand{\thefootnote}{\fnsymbol{footnote}}

\newcommand{\beq}[1]{
\begin{equation}\label{#1}}
\newcommand{\eeq}{\end{equation}}
\newcommand{\bea}[1]{
\begin{eqnarray}\label{#1}}
\newcommand{\eea}{\end{eqnarray}}
\begin{document}
\title{Exploring backward pion electroproduction in the scaling regime}

\author{J.P.~Lansberg$^{a,b}$, B.~Pire$^a$,L. Szymanowski$^{a,b,c}$}{
address={$^a$Centre de Physique Th\'eorique, \'Ecole polytechnique, CNRS, 
91128 Palaiseau, France\\
$^b$Physique th\'eorique fondamentale, Universit\'e de  Li\`ege,  B-4000 Li\`ege~1, Belgium\\
$^{c}$ Soltan Institute  for   Nuclear  Studies,  Warsaw,   Poland \\
E-mail: Jean-Philippe.Lansberg@cpht.polytechnique.fr}
}

\begin{abstract}
We use general relations between the Transition Distribution Amplitudes (TDAs),
entering the description of the $p\to \pi^0$ transition, and the proton Distribution 
Amplitudes (DAs) in the 
soft-pion limit to estimate the size of the amplitude for backward
electroproduction of $\pi^0$ at large $Q^2$. 
\end{abstract}

\maketitle

%%%%%%%%%%%%%%%%%%%%%%%%%%%%%%%%%%%%%%%%%%%%
%% MAINMATTER
%%%%%%%%%%%%%%%%%%%%%%%%%%%%%%%%%%%%%%%%%%%%

\footnotetext{Presented  at {\it Quark Confinement and the Hadron Spectrum VII}, September 2-7 2006, Ponta Delgada, Portugal and
at {\it Soft-Pions in hard processes}, August 3-5 2006, Regensburg, Germany.}
\renewcommand{\thefootnote}{\arabic{footnote}}

%\section{Introduction}

We have recently~\cite{TDApigamma} shown that  factorisation theorems \cite{fact} 
for exclusive processes apply to 
$\pi^-\,\pi^+ \, \to \, \gamma^*\,\gamma$
in the kinematical regime where the virtual photon is highly virtual but at
small  $t$. We also advocated the extension of this approach to 
$P \bar P  \to \gamma ^* \gamma $, to backward VCS 
$  \gamma^\star P  \to P' \gamma $~\cite{Lansberg:2006uh},
to backward pion electroproduction 
$\gamma^\star P  \to P' \pi$ and 
 to 
$ P \bar P \to \gamma^* \pi $
in the  near forward region and for large virtual $Q^2$, which may  be studied in detail  at GSI.

For the  $\gamma^\star$ to $\rho$ transition, a perturbative limit of
the TDA may be obtained~\cite{Pire:2006ik}. For
$ \gamma \to \pi$ one, where there are only four leading-twist 
TDAs~\cite{TDApigamma} related to
$   \langle \gamma|\, {\bar q}_{\alpha}(z_{1} n)\, 
[z_1;z_0]\,{ q}_{\beta}(z_{0} n) \,|\pi \rangle$, where $[z_1;z_0]$ denotes the Wilson line
we have recently shown~\cite{TDApigamma-appl} that experimental analysis of \eg~$\gamma^\star \gamma \to \rho \pi$ 
and  $\gamma^\star \gamma
 \to \pi \pi$ could be carried out since the Bremsstrahlung contribution
is small and rates are sizable at present $e^+ e^-$ facilities.
Whereas in the pion case, models used for GPDs (see~\cite{Bissey:2003yr} and references therein) 
could be applied to TDAs, this is not obvious for baryonic ones, for which the soft limit considered here
is therefore very interesting.

In Ref.~\cite{TDApiproton}, we have defined the leading-twist proton to pion
$P \to \pi$ transition distribution amplitudes from the Fourier transform 
of the matrix element
\begin{equation}\label{eq:mat_el_p-pi}
   \langle \pi|\, \epsilon^{ijk} {q}^i_{\alpha}(z_{1}n)\, 
[z_1;z_0]\,{q}^j_{\beta}(z_{2} n)\, [z_2;z_0]\,
   {q}^k_{\gamma}(z_{3} n)\,[z_3;z_0] \,|P \rangle ,
 \end{equation}
  
We define here the leading-twist TDAs for the $P \to \pi^0$ transition 
at $\Delta_T=0$ as\footnote{In the following,
we shall use the notation $\displaystyle {\cal F}\equiv (p.n)^3\int^{\infty}_{-\infty} dz_i e^{\Sigma_i x_i z_i p.n}$.}:
\bea{TDA}
 && 4 {\cal F}\Big(\langle     \pi^0(p_\pi)|\, \epsilon^{ijk}u^{i}_{\alpha}(z_1 n) 
[z_1;z_0] u^{j}_{\beta}(z_2 n)[z_2;z_0] d^{k}_{\gamma}(z_3 n)[z_3;z_0]
\,|P(p_1,s_1) \rangle \Big)  
\\ \nonumber
&&= i\frac{f_N}{f_\pi}\Big[ V^{p\pi^0}_{1} (\ks p C)_{\alpha\beta}(N^+)_{\gamma}  +
A^{p\pi^0}_{1} (\ks p\gamma^5 C)_{\alpha\beta}(\gamma^5 N^+)_{\gamma} +
\,T^{p\pi^0}_{1} (\sigma_{p\mu} C)_{\alpha\beta}(\gamma^\mu N^+)_{\gamma} \Big], 
\eea
where $\sigma^{\mu\nu}= 1/2[\gamma^\mu, \gamma^\nu]$, $C$ is the charge 
conjugation matrix 
and $N^+$ is the large component of the nucleon spinor ($N=(\ks n \ks p + \ks p \ks n) N = N^-+N^-$
with $N^+\sim \sqrt{p_1^+}$ and $N^-\sim \sqrt{1/p_1^+}$).
 $f_\pi$ is the pion decay constant ($f_\pi = 133$ MeV) and $f_N$ has been estimated through 
QCD sum rules to be of
order $5.2\cdot 10^{-3}$ GeV$^2$~\cite{CZ}. All the TDAs $V_i$, $A_i$ and $T_i$ are
dimensionless.

Now, we shall derive the general limit
of these three contributing TDAs at $\Delta_T=0$  when $\xi$ gets close
to 1. In that limit, the soft-meson theorems~\cite{AD} derived from current algebra
apply~\cite{PPS}, which allow us to express these 3 TDAs in terms  of the 3 Distribution
Amplitudes (DAs) of the corresponding baryon. In the case of the proton DA~\cite{CZ}, 
$V^p(x_i)$, $A^p(x_i)$, $T^p(x_i)$ are defined such as 
\bea{eq:DA}
&&4{\cal F}\Big(\langle 0|\epsilon_{ijk}u^i_\alpha(z_1 n)u^j_\beta(z_2 n)d^k_\gamma(z_3 n)|p(p,s)\rangle\Big)
  = f_N \times \\&&\Big[
V^p(x_{i}) ( \ks p C)_{\alpha \beta} (\gamma^5 N^+)_\gamma 
+ A^p(x_{i}) (\ks p \gamma^5 C)_{\alpha \beta} N^+_\gamma
+ T^p(x_{i}) (\sigma_{p  \mu}\,C)_{\alpha \beta} 
(\gamma^\mu \gamma^5 N^+)_\gamma 
\Big].\nonumber
\eea

We use the general soft pion 
theorem~\cite{AD} to write:
\begin{eqnarray}
\langle \pi^a(p_\pi)  |{\cal O}| P(p_1,s_1)\rangle =-\frac{i}{f_\pi} \langle 0  | [ Q^a_5,  {\cal O}]  | P(p_1,s_1) \rangle +\hbox{ pole term} 
\label{eq:soft-theorem}
\end{eqnarray}

The second term, which takes care of the nucleon pole term, does not contribute at threshold and
will not be considered in the following.

For the transition $P\to \pi^0$, $Q^a_5=Q^3_5$ and the flavour content of ${\cal O}$ is $u_\alpha u_\beta d_\gamma$. 
Since the commutator of the chiral charge $Q_{5}$ with the quark field $\psi$ ($\tau^a $ being the isospin matrix) is
$[Q_{5}^a, \psi] = - \frac{\tau^a}{2} \gamma^5 \psi\;$,
the first term in the rhs of \ce{eq:soft-theorem} gives three terms from 
$(\ga u)_\alpha u_\beta d_\gamma$, $u_\alpha (\ga u)_\beta d_\gamma$ and $u_\alpha u_\beta (\ga d)_\gamma$.
The corresponding multiplication by $\ga$ (or $(\ga)^T$ when it acts on the index $\beta$) on the 
vector and axial structures of the DA (\ce{eq:DA}) gives two terms which cancel each other 
and the third one,
which remains, is the same as the one for the TDA, up to the modification that in the DA decomposition
$p$ is the proton momentum, whereas for the TDA one, $p$ is the light-cone projection of $P\equiv (p_1+p_\pi)/2$, $\ie$ half
the proton momentum if one neglects $p_\pi$. This introduces a factor 2 in the relation
between the DA $V^p$ ($V^p$) and the TDA $V^{p\pi^0}_{1}$ ($A^{p\pi^0}_{1}$), which  
cancels the factor $1/2$ from $[Q_{5}^a, \psi]$. 
To what concerns the tensorial structure multiplying $T^p$,  the three 
terms are identical at leading-twist accuracy and yield a factor 3 in $T_1$.

We eventually have the soft limit\footnote{The factor 
$\frac{1}{2}$ in the argument of the DA in~\ce{eq:softp} comes from the fact that for the TDAs, 
the $x_i$ are defined with respect to $p$ 
( see \eg~$ {\cal F}\equiv (p.n)^3\int^{\infty}_{-\infty} dz_i e^{\Sigma_i x_i z_i p.n}$) and
  $p  \to \frac{p_1}{2}$ when $\xi \to 1$. Therefore, they
vary within the interval $[0:2]$, whereas for the DAs, the momentum fraction are defined with
respect to the proton momentum $p_1$ and vary between 0 and 1.}
for our three TDAs at $\Delta_T=0$:
\eqs{\label{eq:softp}
V^{p\pi^0}_1(x_i,\xi,t) \to  V^p \Big(\frac{x_i}{2}\Big),~
A^{p\pi^0}_1(x_i,\xi,t) \to A^p \Big(\frac{x_i}{2}\Big), ~
T^{p\pi^0}_1(x_i,\xi,t) \to 3 T^p \Big(\frac{x_i}{2}\Big) 
.}

At  leading order in $\alpha_s$, the amplitude  for 
 $\gamma^\star(q) P(p_1,s_1) \to P'(p_2,s_2) \pi^0(p_\pi)$ is
\begin{eqnarray}
{\cal M}^\mu =  -i e F^{p\pi^0}(Q^2,\xi,t)  \bar u(p_2) \gamma^\mu \gamma^5 u(p_1),
F^{p\pi^0}  =\frac{C f_{N}^2}{f_{\pi}Q^4}
 \int\limits^{1+\xi}_{-1+\xi} \! \! \! d^3x \int\limits_0^1 \! \!d^3y
\sum\limits_{\alpha=1}^{14} T'_{\alpha}(x_{i},y_{j}),
\end{eqnarray}
to be compared with the leading amplitude for the baryonic form factor~\cite{CZ}
\begin{eqnarray}
  {\cal M}^\mu =  -i e F_1^{p}(Q^2) \bar u(p_2) \gamma^\mu u(p_1),
F_1^{p}  =\frac{C f_{N}^2}{Q^4}  \int\limits_{0}^{1} d^3x \int\limits_0^1 d^3y
\sum\limits_{\alpha=1}^{14} T_{\alpha}(x_{i},y_{j}).
\end{eqnarray}

Considering, for now, only the contribution from the ERBL region $x_i>0$, the integration
between $-1+\xi$ and ${1+\xi}$ can be converted into one between 0 and 1 by a change of variable. 
Since
the expressions of $T'_\alpha$ and $T_\alpha$ are identical up to the 3 replacements
the initial-state DAs by the $P\to \pi^0$ TDAs, they would in 
fact  differ only by the factor 3 in the last relation of~\ce{eq:softp} 
extrapolating the $\xi \to 1$ limit to the ERBL region,.

Due to this factor, whereas the asymptotic choice~\cite{CZ}, $120~x_1 x_2 x_3$, for the DAs gives a
vanishing result for $F_1^{p}$ or $G^p_M$, the result is nonzero for $F^{p\pi^0}$. This lets  us 
therefore hope
that the onset of the dominance of the perturbative contribution to $\gamma^\star P \to P' \pi^0$
 with TDAs may happen at much lower $Q^2$ than for the proton form factor. 
Quantitative results to be compared with the measurement of~\cite{Laveissiere:2003jf} 
will be presented soon.

%%%%%%%%%%%%%%%%%%%%%%%%%%%%%%%%%%%%%%%%%%%%%%%%
%% BACKMATTER
%%%%%%%%%%%%%%%%%%%%%%%%%%%%%%%%%%%%%%%%%%%%%%%%

\begin{theacknowledgments}
This  work  is  partially supported  by  the
scientific agreement Polonium, the Polish Grant 1 P03B 028 28, the EU program I3HP,
 contract RII3-CT-2004-506078 and the FNRS (Belgium). 
L.Sz. is a Visiting Fellow of the FNRS (Belgium).
\end{theacknowledgments}

\bibliographystyle{aipprocl} % if natbib is missing

\end{document}